\newcommand{\be}{\begin{equation}}  
\newcommand{\ee}{\end{equation}}  
\newcommand{\beq}{\begin{eqnarray}}  
\newcommand{\eeq}{\end{eqnarray}}
\documentclass{PoS}

\usepackage{amsmath}
\usepackage{graphicx}
\usepackage{mathtools}

\title{Computation of correlation matrices for tetraquark candidates with $J^P = 0^+$ and flavor structure $q_1 \bar{q_2} q_3 \bar{q}_3$}

\ShortTitle{Computation of correlation matrices for tetraquark candidates with $J^P = 0^+$}

\author{\speaker{Joshua Berlin}${}^{1}$, Abdou Abdel-Rehim${}^{2}$, Constantia Alexandrou${}^{2,3}$, Mattia Dalla Brida${}^{4}$, Mario Gravina${}^{5}$, Marc Wagner${}^{1}$\\

\email{berlin@th.physik.uni-frankfurt.de},
\email{a.abdel-Rehim@cyi.ac.cy},
\email{alexand@ucy.ac.cy},
\email{mattia.dalla.brida@desy.de},
\email{mario.gravina@fis.unical.it},
\email{mwagner@th.physik.uni-frankfurt.de}\\

        ${}^{1}$Goethe-Universit\"at Frankfurt am Main, Institut f\"ur theoretische Physik, \\ $\quad$ Max-von-Laue-Stra{\ss}e 1, D-60438 Frankfurt am Main, Germany\\
        ${}^{2}$Computation-based Science and Technology Research Center, The Cyprus Institute, \\ $\quad$ 20 Kavafi Street, 2121 Nicosia, Cyprus\\
        ${}^{3}$Department of Physics, University of Cyprus, P.O.\ Box 20537, 1678 Nicosia, Cyprus\\
        ${}^{4}$NIC, DESY, Platanenallee 6, 15738 Zeuthen, Germany\\
        ${}^{5}$Universit\`a della Calabria, Via Pietro Bucci, 87036 Arcavacata di Rende Cosenza, Italy \\
        }

\abstract{
The conjecture that several recently observed mesons have a structure, which is not dominated by an ordinary quark-antiquark pair, but by a four-quark structure, is being actively investigated both theoretical and experimentally.  Such a state may be characterized as a mesonic molecule or as a diquark-antidiquark pair. Lattice QCD provides a theoretically sound framework to study such states. To quantitatively investigate the internal structure of such mesons, one needs to precisely compute correlation matrices containing several interpolating operators including two and four quarks. Here we discuss certain technical aspects of such correlation matrices suited to study tetraquark candidates with $J^P = 0^+$ and flavor structure $q_1 \bar{q_2} q_3 \bar{q}_3$, e.g.\ the $a_0(980)$ meson, the $D_{s0}^\ast$ meson and some of the charged $c \bar{c}$ $X$ states. Some numerical results for the $a_0(980)$ meson are presented.
}

\FullConference{The 33rd International Symposium on Lattice Field Theory\\
		 14-18 July, 2015\\
		 Kobe International Conference Center, Japan}

\begin{document}


\section{\label{sec:mot}Motivation}

\begin{figure}[t]
\begin{center}
\includegraphics[width=4.95cm,page=2]{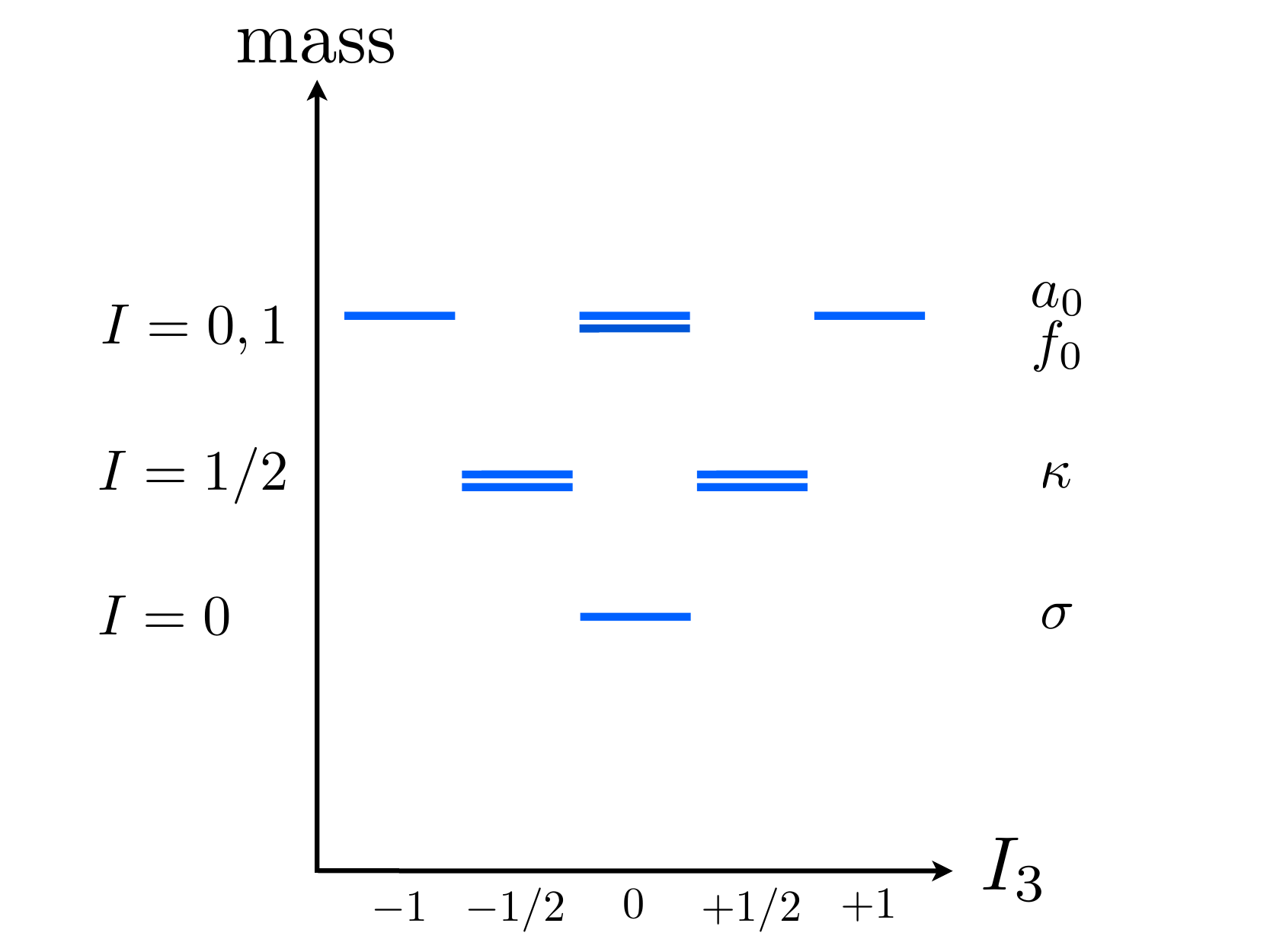}
\includegraphics[width=4.95cm,page=4]{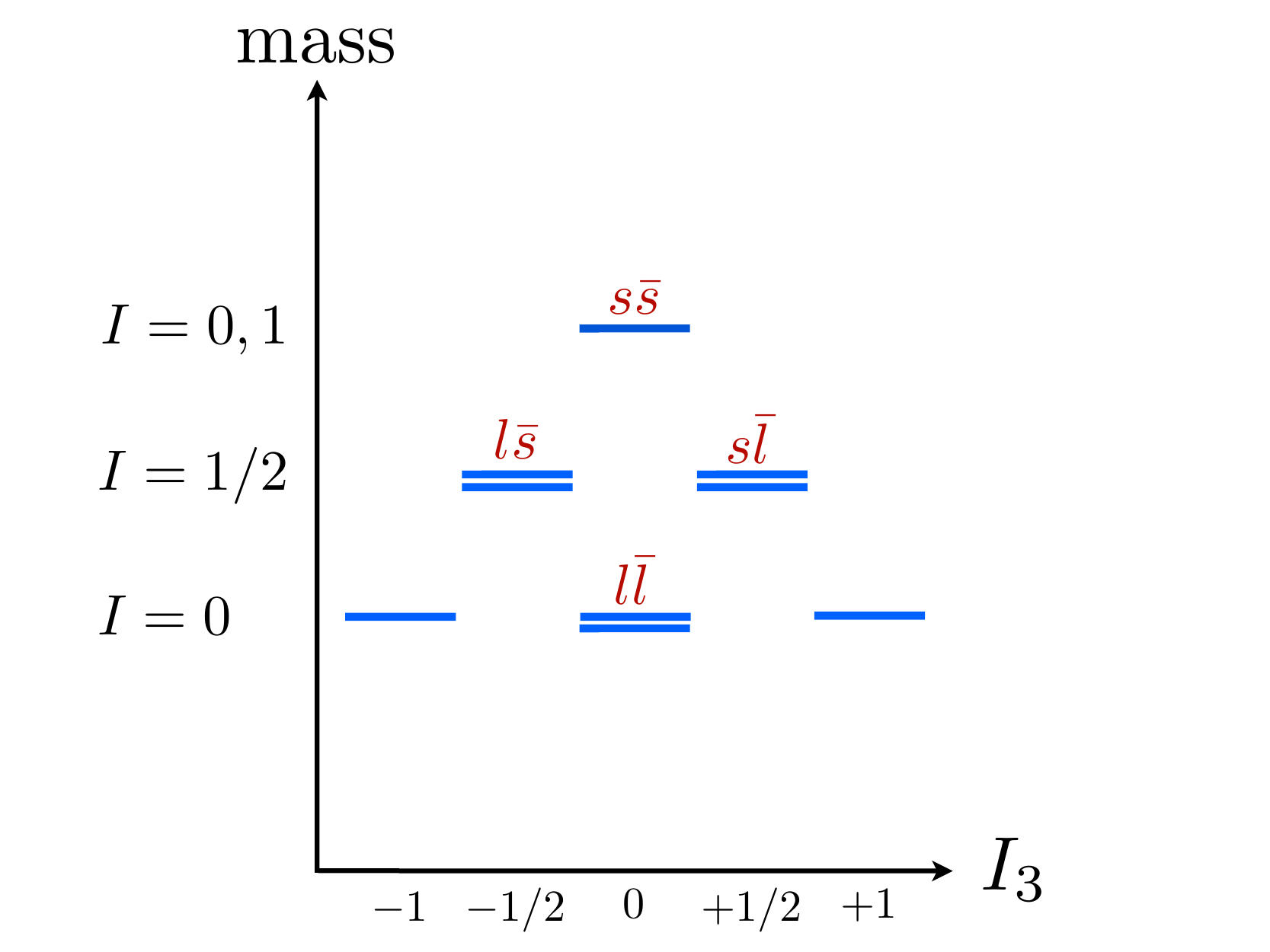}
\includegraphics[width=4.95cm,page=3]{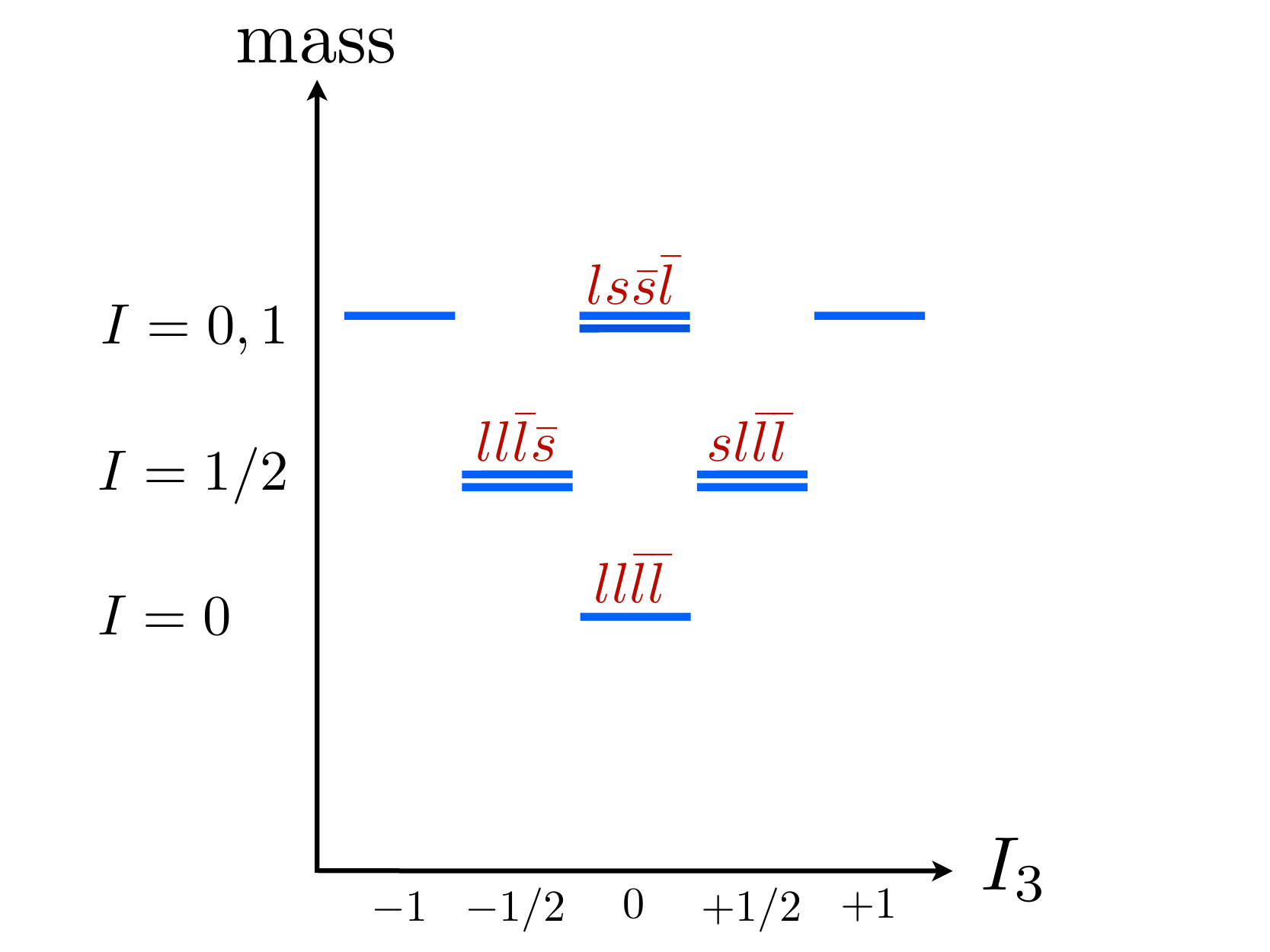}
\end{center}
\caption{\label{fig:lightscalarspectrum}Spectrum of light scalar mesons $(J^P = 0^+)$. Experimental results ~\textbf{(left)}; Theoretical expectation within the conventional quark model ($q \bar q$ structure) ~\textbf{(center)}; Theoretical expectation assuming a four-quark structure ~\textbf{(right)}.}
\end{figure}

Our understanding of the light scalar meson sector (quantum numbers $J^P = 0^+$) is still incomplete \cite{Pelaez:2013jp,Pelaez:2014rla,pdg:sep2014}. The observed mass ordering of $\sigma$, $\kappa$, $f_0(980)$ and $a_0(980)$ is inverted compared to expectations from conventional $q \bar q$ quark model calculations. Moreover, when assuming a $q \bar q$ structure, it is hard to explain the mass degeneracy of $f_0(980)$ and $a_0(980)$. However, the interpretation of this meson sector assuming a four-quark structure is rather straightforward. The expected mass ordering is consistent with experimental results and due to an additional $s \bar s$ pair the degeneracy of $f_0(980)$ and $a_0(980)$ is not surprising (cf.\ Figure~\ref{fig:lightscalarspectrum}).

Several lattice QCD studies of the light scalar mesons have been published in the last couple of years \cite{Bernard:2007qf,Gattringer:2008be,Prelovsek:2008qu,Liu:2008ee,Wakayama:2012ne,
Prelovsek:2013ela,Wakayama:2014gpa}. In this work we continue our investigation of the $a_0(980)$ meson \cite{Daldrop:2012sr,Alexandrou:2012rm,Wagner:2012ay,
Wagner:2013nta,Wagner:2013jda,Wagner:2013vaa,Abdel-Rehim:2014zwa} and briefly discuss the applicability of our methods and codes to other systems, the $D_{s0}^\ast$ meson and some of the charged $c \bar{c}$ $X$ states. In particular the investigation of tetraquark candidates with two heavy
and two light quarks seems promising, since recent computations in the
static limit predict a bound state \cite{Bicudo:2015vta,Peters:2015tra}.


\section{\label{sec:operators}Interpolating operators and the correlation matrix}

Our investigations are based on a $6 \times 6$ correlation matrix
\begin{align}
\label{EQN100} C_{j k}(t) = \Big\langle \mathcal{O}^j(t_2) \mathcal{O}^{k \dag}(t_1) \Big\rangle \quad , \quad t = t_2 - t_1 .
\end{align}
The interpolating operators $\mathcal{O}^j$ generate quantum numbers $I(J^P) = 1(0^+)$,
\begin{align}
\mathcal{O}^1 = \mathcal{O}^{q\bar q} =& \sum_{\bf{x}} \Big( {\bar d}_{\bf x} {u}_{\bf x} \Big) \label{eq:operatorone}
 \\
\mathcal{O}^2 = \mathcal{O}^{K \bar{K} \text{, point}} =& \sum_{\bf{x}} \Big( {\bar s}_{\bf x} \gamma_5 {u}_{\bf x} \Big) \Big( {\bar d}_{\bf x} \gamma_5 {s}_{\bf x} \Big) \label{eq:operatortwo}
\\
\mathcal{O}^3 = \mathcal{O}^{\eta_s \pi \text{, point}} =& \sum_{\bf{x}} \Big( {\bar s}_{{\bf x}} \gamma_5 {s}_{{\bf x}} \Big) \Big( {\bar d}_{{\bf x}} \gamma_5 {u}_{{\bf x}} \Big) \label{eq:operatorthree}
 \\
\mathcal{O}^4 = \mathcal{O}^{Q \bar Q} =& \sum_{\bf{x}} \epsilon_{abc} \Big( {\bar s}_{{\bf x}, b} {(C \gamma_5)} {\bar d}^T_{{\bf x}, c} \Big) \epsilon_{ade} \Big( {u}^T_{{\bf x}, d} {(C \gamma_5)} s_{{\bf x}, e} \Big) \label{eq:operatorfour}
\\
\mathcal{O}^5 = \mathcal{O}^{K\bar{K} \text{, 2-part}} =& \sum_{{\bf x,y}} \Big( {\bar s}_{{\bf x}} \gamma_5 {u}_{{\bf x}} \Big) \Big( {\bar d}_{{\bf y}} \gamma_5 {s}_{{\bf y}} \Big) \label{eq:operatorfive}
\\
\mathcal{O}^6 = \mathcal{O}^{\eta_s \pi \text{, 2-part}} =& \sum_{\bf{x,y}} \Big( {\bar s}_{{\bf x}} \gamma_5 {s}_{{\bf x}} \Big) \Big( {\bar d}_{{\bf y}} \gamma_5 {u}_{{\bf y}} \Big) , \label{eq:operatorsix}
\end{align}
where $C$ is the charge conjugation matrix. The operator $\mathcal{O}^{q\bar q}$ generates a standard quark-antiquark state, while all other operators generate four-quark states. $\mathcal{O}^{K \bar{K} \text{, point}}$ and $\mathcal{O}^{\eta_s \pi \text{, point}}$ are of mesonic molecule structure ($K \bar{K}$ and $\eta_s \pi$), while $\mathcal{O}^{Q \bar Q}$ corresponds to a diquark-antidiquark pair (we use the lightest (anti)diquarks with spin structure $C \gamma_5$ \cite{Jaffe:2004ph,Alexandrou:2006cq,Wagner:2011fs}). These three operators are intended to model the expected structures of possibly existing four-quark bound states, i.e.\ of tetraquarks. The remaining two operators $\mathcal{O}^{K\bar{K} \text{, 2-part}}$ and $\mathcal{O}^{\eta_s \pi \text{, 2-part}}$ independently generate two mesons ($K + \bar{K}$ and $\eta_s + \pi$) and, hence, should be suited to resolve low-lying two-particle scattering states.

\begin{figure}[htb]
\begin{center}
\includegraphics[width=15.cm,page=1]{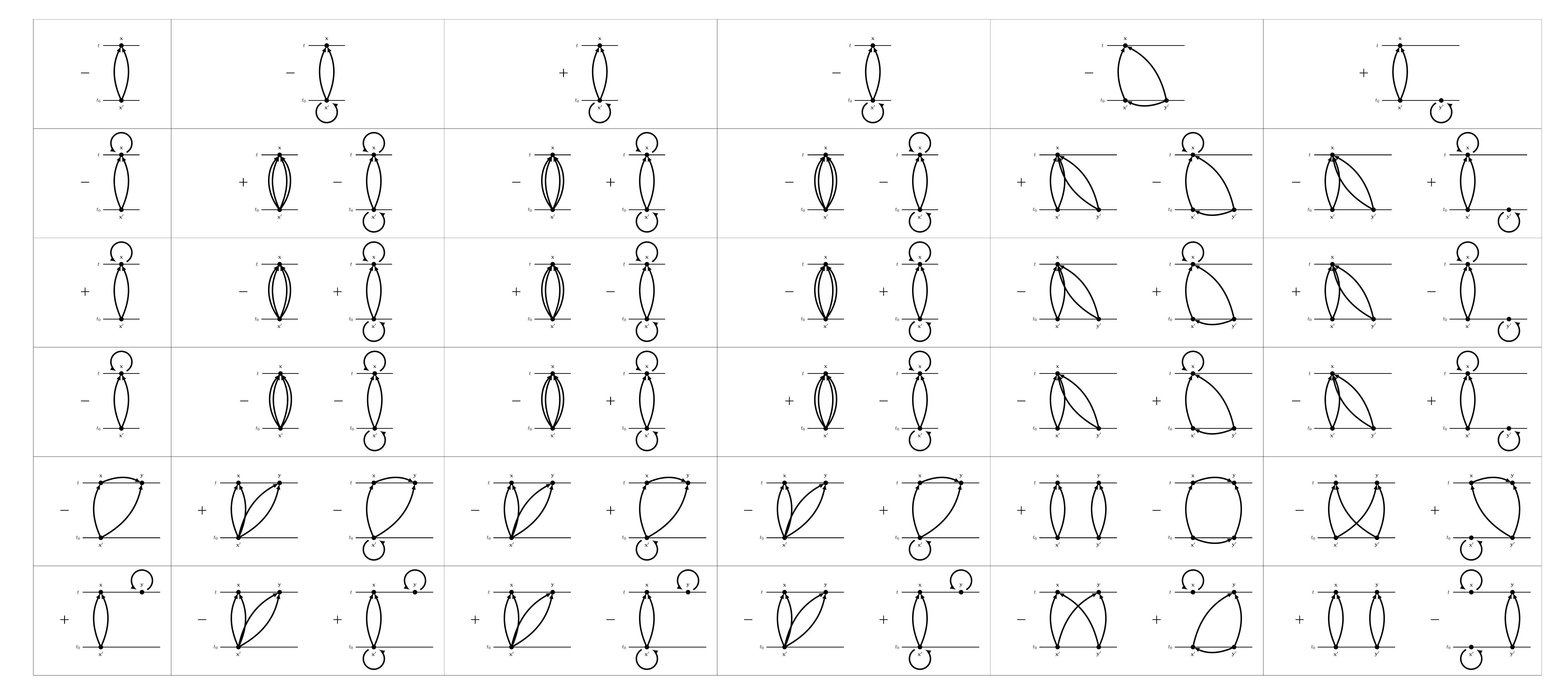}
\end{center}
\caption{\label{fig:diagrammatrix}Diagrammatic representation of the $6 \times 6$ correlation matrix $C_{j k}$ containing the interpolating operators \protect\eqref{eq:operatorone} to \protect\eqref{eq:operatorsix}. The diagrams only reflect the spatial quark propagation. Even though elements with $j,k \in \{ 2, 3, 4 \}$ appear to be identical, they differ in their spin and color structures.}
\end{figure}

In Figure~\ref{fig:diagrammatrix} the correlation matrix $C_{j k}$ is shown in a graphical way in terms of diagrams, where each line represents a quark propagator. Clearly there are disconnected diagrams and diagrams, where quarks propagate within a timeslice. Computing these diagrams precisely is very challenging (cf.\ section~\ref{sec:techniques} and \cite{Abdel-Rehim:2014zwa}).

Note that the correlation matrix given in Eq. (\ref{EQN100}) is also suited to studying the $D_{s0}^\ast$ meson, when replacing the quark flavors according to $u \rightarrow s$, $d \rightarrow c$ and $s \rightarrow u, d$. Similarly, one can explore certain charged charmed tetraquarks, when replacing $s \rightarrow c$.


\section{\label{sec:techniques} Techniques to compute the correlation matrix elements}

A number of diagrams of the correlation matrix shown in Figure~\ref{fig:diagrammatrix} are challenging to compute, in particular disconnected diagrams or diagrams, where quarks propagate within a timeslice. Our strategy is to combine several standard techniques including (a)~fixed-source propagators, (b)~stochastic timeslice-to-all propagators, (c)~the one-end trick and (d)~sequential propagators (cf.\ \cite{Abdel-Rehim:2014zwa} and references therein). For most diagrams there are a couple of possibilities, where in many cases the most efficient combination with respect to the required HPC resources and the obtained statistical accuracy can only be identified after extensive numerical tests and/or analytical considerations. Examples dealing with $C_{1 1}$ and the disconnected diagram of $C_{4 6}$ can be found in \cite{Abdel-Rehim:2014zwa}. Below we briefly discuss another example, $C_{1 5}$. A complete and detailed discussion of the full correlation matrix will be part of an upcoming publication.

\begin{figure}[htb]
\begin{center}
\begin{minipage}{0.48\textwidth} 
\hspace{-1cm}\includegraphics[width=9cm,page=1]{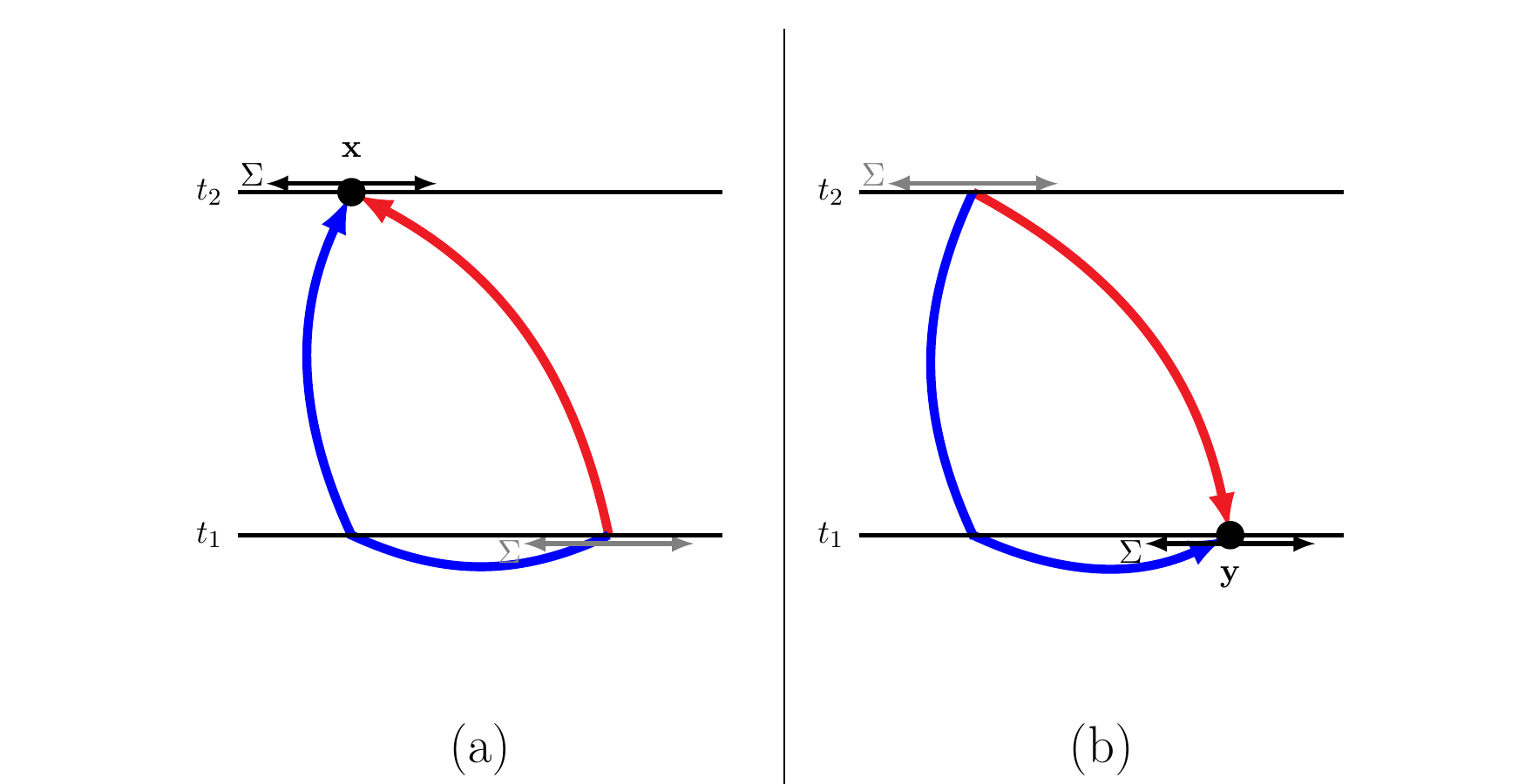}\vspace{.5cm}
\end{minipage}
\begin{minipage}{0.48\textwidth} 
\includegraphics[width=7.5cm,page=3]{pos_relevant}
\end{minipage}
\end{center}
\vspace{-0.5cm}
\caption{\label{fig:triangulardiagram}Two possibilities to compute $C_{1 5}$ using the one-end trick and sequential propagators. Both methods show comparable errors, but method (b) requires with $t/a+1$ inversions a more computational effort.}
\end{figure}

The matrix element
\begin{eqnarray}
\label{eq:correxample} C_{1 5}(t) = -\bigg\langle \sum_{\mathbf{x},\mathbf{y},\mathbf{z}} \mathrm{tr}\Big(G^s({\bf z},t_1;{\bf y},t_1)^\dag \gamma_5 G^u({\bf x},t_2;{\bf z},t_1)^\dag\gamma_5 G^d({\bf x},t_2;{\bf y},t_1)\Big) \bigg\rangle
\end{eqnarray}
(quark propagators are denoted by $G^q$) can e.g.\ be computed by combining the one-end trick and the technique of sequential inversions. The one-end trick realizes an implicit summation over a timeslice, which can be either at $t_1$ (case~(a)) or at $t_2$ (case (b)),
\begin{eqnarray}
\hspace{-0.7cm} \text{(a):}& 
- \langle \sum_{\bf x,z} \mathrm{tr}\Big(
\underbracket{\textcolor{blue}{[\phi^s(z) \delta_{z_0,t_1}]^\dag \gamma_5 G^u({\bf x}, t_2;{\bf z},t_1)^\dag}}_\text{seq. inversion} \gamma_5 \textcolor{red}{\phi^d(x)}
\Big)\Big\rangle
\hspace{.2cm} &\rightarrow 
- \Big\langle \sum_{\bf x} \mathrm{tr}\Big(
\textcolor{blue}{\psi^{u/s}(x)^\dag} \gamma_5 \textcolor{red}{\phi^d(x)}
\Big)\Big\rangle 
\\
\hspace{-0.7cm} \text{(b):}& 
- \Big\langle \sum_{\bf y,z} \mathrm{tr}\Big(
\textcolor{red}{\phi^d(y)^\dag} \gamma_5 \underbracket{\textcolor{blue}{G^s({\bf y}, t_1;{\bf z},t_1) \gamma_5  \phi^u(z)} }_\text{seq. inversion} 
\Big)\Big\rangle
&\rightarrow 
- \Big\langle \sum_{\bf y} \mathrm{tr}\Big(
\textcolor{red}{\phi^{d}(y)^\dag} \gamma_5 \textcolor{blue}{\psi^{s/u}(y)}
\Big)\Big\rangle  
\end{eqnarray}
(cf.\ also Figure~\ref{fig:triangulardiagram}). At first glance both options seem to be equivalent. Note, however, that the computation of the sequential propagator (the blue lines in Figure~\ref{fig:triangulardiagram}), requires for case~(a) only $2$ inversions of the Dirac matrix, while for case~(b) $t_\textrm{max}/a + 1$ inversions are necessary to obtain $C_{1 5}(t)$ for $1 \leq t \leq t_\textrm{max}$. Clearly, proceeding as in (a) is more efficient than proceeding as in (b).

\section{\label{sec:results}Numerical results}

\begin{figure}[h!]
\begin{center}
\includegraphics[width=14cm,page=4]{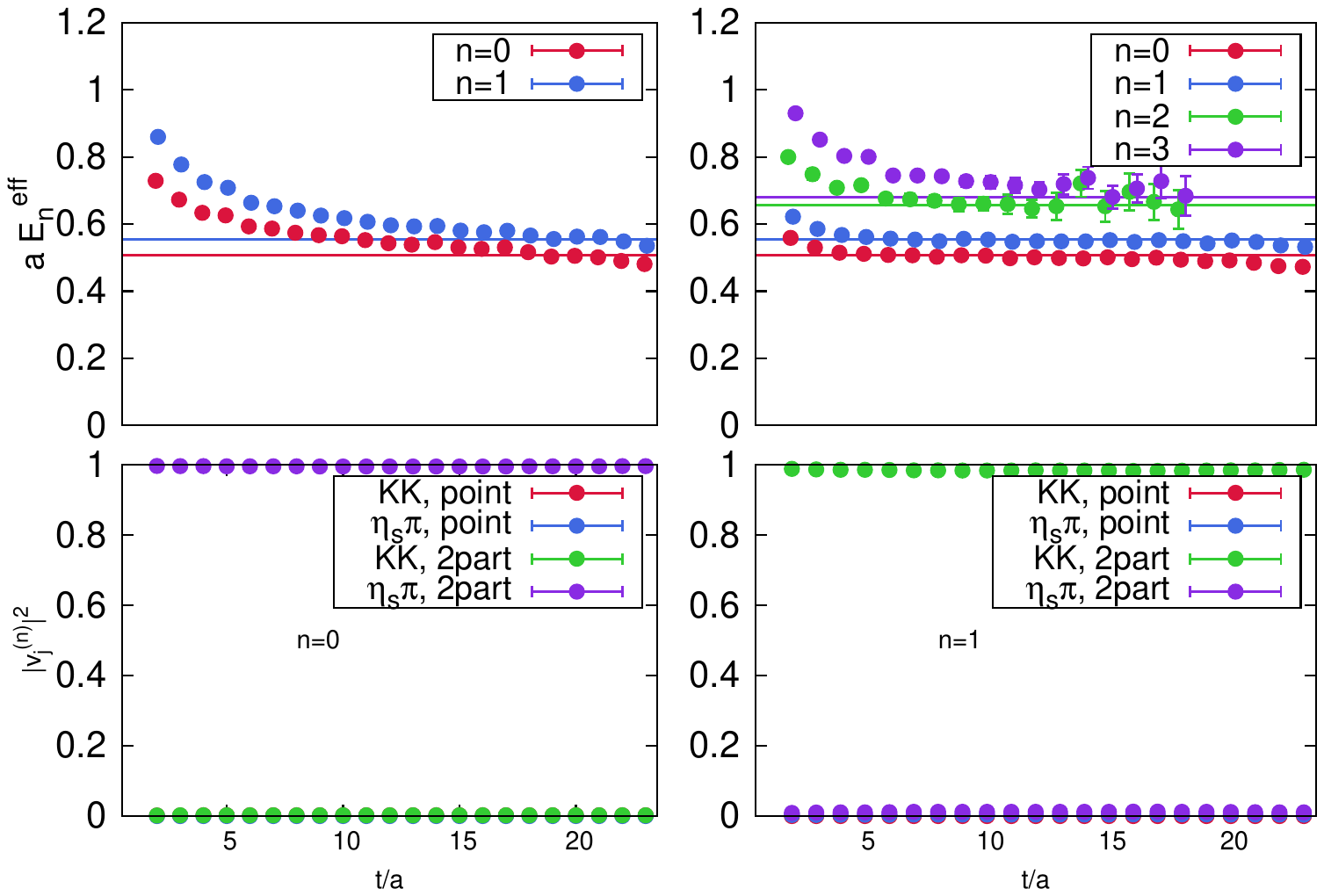}\\
\vspace{-2.8cm}
\hspace{-2.7cm}\includegraphics[width=14cm,page=5]{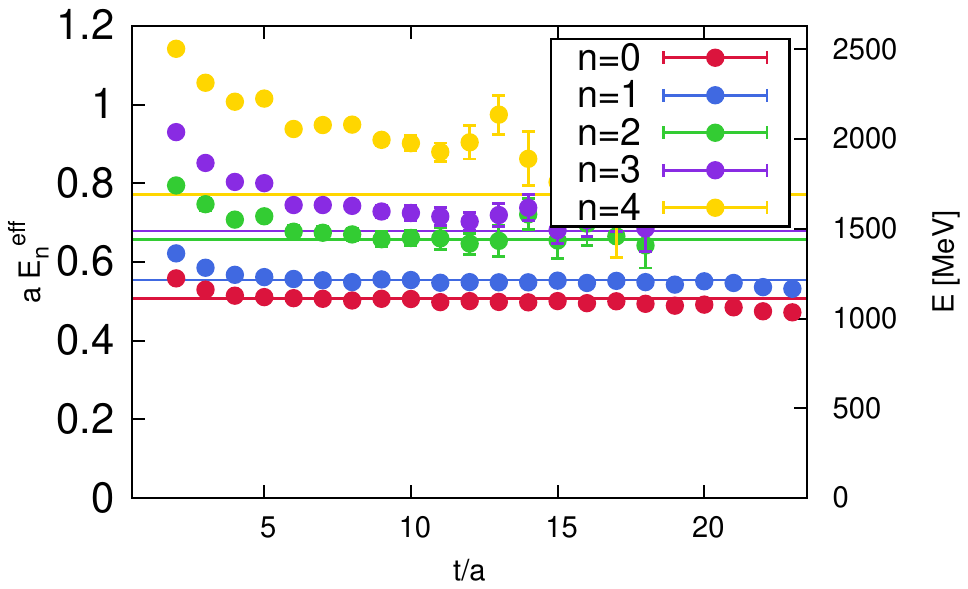}
\end{center}
\vspace{-4.cm}
\caption{\label{fig:results}\textbf{(top left)}~Effective masses from the $2\times2$ correlation matrix containing $\mathcal{O}^{K\bar K\text{,point}}$ and $\mathcal{O}^{\eta_s \pi \text{,point}}$. \textbf{(top right)}~Effective masses from the $4\times4$ correlation matrix containing additionally $\mathcal{O}^{K\bar K\text{,2-part}}$ and $\mathcal{O}^{\eta_s \pi\text{,2-part}}$. (The horizontal lines correspond to expected energies of the two-meson states $K + \bar{K}$ and $\eta_s + \pi$, where both mesons are at rest and with one relative quantum of momentum.) \textbf{(center left and right)}~Squared eigenvector components obtained from the $4 \times 4$ correlation matrix. \textbf{(bottom)}~Effective masses from the $5 \times 5$ correlation matrix containing additionally $\mathcal{O}^{Q \bar Q}$.}
\end{figure}

At the moment we are performing computations as described in section~\ref{sec:operators}, section~\ref{sec:techniques} and Ref.~\cite{Abdel-Rehim:2014zwa} using around 500 Wilson clover gauge link configurations with 2+1 dynamical quark flavors generated by the PACS-CS collaboration \cite{Aoki:2008sm}. The lattice size is $32^3 \times 64$, the lattice spacing $a \approx 0.09 \, \textrm{fm}$ and the light $u/d$ quark mass corresponds to $m_\pi \approx 300 \, \textrm{MeV}$. Since computations are ongoing and statistical accuracy is currently rather limited, we neglect in a first analysis diagrams of the correlation matrix, where quarks propagate within a timeslice, e.g.\ the diagram shown in Figure~\ref{fig:triangulardiagram} and all diagrams with closed quark loops. Consequently, $C_{1 j} = C_{j 1} = 0$ for $j = 2, \ldots , 6$, i.e.\ the correlation matrix shown in Figure~\ref{fig:diagrammatrix} is reduced to a $5 \times 5$ matrix containing the interpolating operators $\mathcal{O}^2 , \ldots , \mathcal{O}^6$.

We determine effective masses $E_n^\textrm{eff}$ and corresponding eigenvectors $\mathbf{v}_n$ by solving the standard generalized eigenvalue problem
\begin{eqnarray}
C(t) \mathbf{v}_n(t,t_r) = \lambda_n(t,t_r) C(t_r) \mathbf{v}_n(t,t_r) \quad , \quad E_n \overset{t \textrm{ large}}{=} E_n^\textrm{eff}(t,t_r) = \frac{1}{a} \ln\bigg(\frac{\lambda_n(t,t_r)}{\lambda_n(t+a, t_r)}\bigg)
\end{eqnarray}
with $t_r = a$.

Effective masses from the $2 \times 2$ correlation matrix containing only the mesonic molecule operators $\mathcal{O}^{K\bar K\text{,point}}$ and $\mathcal{O}^{\eta_s \pi \text{,point}}$ are shown in Figure~\ref{fig:results} (top left). There are two states at around $1 \, \textrm{GeV}$, i.e.\ roughly consistent with the expected mass of the $a_0(980)$ meson \cite{PDG} as well as with the two-meson states $K + \bar{K}$ and $\eta_s + \pi$ in our lattice setup.

Similarly, effective masses from the $4 \times 4$ correlation matrix containing additionally the two-meson operators $\mathcal{O}^{K\bar K\text{,2-part}}$ and $\mathcal{O}^{\eta_s \pi\text{,2-part}}$ are shown in Figure~\ref{fig:results} (top right). The same two low-lying states around $1 \, \textrm{GeV}$ are obtained, but the corresponding plateaus are reached at smaller temporal separations $t$ and exhibit less statistical fluctuations. The energies of the second and third excitations are significantly larger, consistent with the expected energies of two-meson states with one relative quantum of momentum. This indicates that the two low-lying states are of two-meson type, i.e.\ neither of the two corresponds to a bound four-quark state, which can be identified with the $a_0(980)$ meson. This is further supported by the squared eigenvector components of $\mathbf{v}_0$ and $\mathbf{v}_1$ obtained from the $4 \times 4$ correlation matrix, which are also shown in Figure~\ref{fig:results} (center left and right).

Results of an analogous analysis of a $5 \times 5$ correlation matrix containing also the diquark-antidiquark operator $\mathcal{O}^{Q \bar Q}$ are shown in Figure~\ref{fig:results} (bottom). Since no additional state at around $1 \, \textrm{GeV}$ appears, the conclusions are essentially the same as for the $4 \times 4$ correlation matrix: \textit{It thus seems that when neglecting quark propagation within a timeslice, and given the unphysical value of the pion mass we are investigating, the $a_0(980)$ meson is not a rather stable four-quark state of mesonic molecule or diquark-antidiquark type.} (Cf.\ also \cite{Alexandrou:2012rm} for a more detailed very similar discussion.)


\begin{acknowledgments}

M.W.\ and J.B.\ acknowledge support by the Emmy Noether Programme of the DFG (German Research Foundation), grant WA 3000/1-1.
The work of M.G.\ was supported by the European Commission, European Social Fund and Calabria Region, that disclaim any liability for the use that can be done of the information provided in this paper.

This work was supported in part by the Helmholtz International Center for FAIR within the framework of the LOEWE program launched by the State of Hesse.

Computations have been performed using the Chroma software library \cite{Edwards:2004sx}. Calculations on the LOEWE-CSC high-performance computer of Johann Wolfgang Goethe-University Frankfurt am Main were conducted for this research. We would like to thank HPC-Hessen, funded by the State Ministry of Higher Education, Research and the Arts, for programming advice.

\end{acknowledgments}



\end{document}